\documentclass{emulateapj}
\usepackage{natbib}\usepackage{times}
\usepackage[usenames,dvipsnames]{xcolor}
\usepackage{bigstrut}

\begin{document}

\shortauthors{Horiuchi et al.}
\title{Effects of stellar rotation on star formation rates and comparison to core-collapse supernova rates}
\shorttitle{Star formation and supernova rates with stellar rotation}
\author{Shunsaku Horiuchi\altaffilmark{1,2}, John F.~Beacom\altaffilmark{2,3,4}, Matt~S.~Bothwell\altaffilmark{5,6}, Todd A.~Thompson\altaffilmark{3,4}} 
\altaffiltext{1}{Center for Cosmology, University of California Irvine, 4129 Frederick Reines Hall, University of California, Irvine, CA 92697; s.horiuchi@uci.edu, http://www.physics.uci.edu/}
\altaffiltext{2}{Center for Cosmology and Astro-Particle Physics, The Ohio State University, 191 West Woodruff Ave., Columbus, OH 43210, http://ccapp.osu.edu/}
\altaffiltext{3}{Department of Physics, The Ohio State University, 191 W.\ Woodruff Ave., Columbus, OH 43210, http://www.physics.ohio-state.edu/}
\altaffiltext{4}{Department of Astronomy, The Ohio State University, 140 West 18th Ave., Columbus, OH 43210, http://www.astronomy.ohio-state.edu/}
\altaffiltext{5}{Steward Observatory, University of Arizona, Tucson, AZ 85721, USA, http://www.as.arizona.edu/}
\altaffiltext{6}{Cavendish Laboratory, University of Cambridge, 19 J.J. Thomson Avenue, Cambridge, CB3 0HE, UK, http://www.phy.cam.ac.uk/}

\begin{abstract}
We investigate star formation rate (SFR) calibrations in light of recent developments in the modeling of stellar rotation. Using new published non-rotating and rotating stellar tracks, we study the integrated properties of synthetic stellar populations and find that the UV to SFR calibration for the rotating stellar population is $30$\% smaller than for the non-rotating stellar population, and $40$\% smaller for the H$\alpha$ to SFR calibration. These reductions translate to smaller SFR estimates made from observed UV and H$\alpha$ luminosities. Using the UV and H$\alpha$ fluxes of a sample of $\sim 300$ local galaxies, we derive a total (i.e., sky-coverage corrected) SFR within $11$ Mpc of $120$--$170 \, {\rm M_\odot \, yr^{-1}}$ and $80$--$130 \, {\rm M_\odot \, yr^{-1}}$ for the non-rotating and rotating estimators, respectively. Independently, the number of core-collapse supernovae discovered in the same volume requires a total SFR of $270^{+110}_{-80} \, {\rm M_\odot \, yr^{-1}}$, suggesting a tension with the SFR estimates made with rotating calibrations. More generally, when compared with the directly estimated SFR, the local supernova discoveries strongly constrain any physical effects that might increase the energy output of massive stars, including, but not limited to, stellar rotation. The cosmic SFR and cosmic supernova rate data on the other hand show the opposite trend, with the cosmic SFR higher than that inferred from cosmic supernovae, constraining a significant decrease in the energy output of massive stars. Together, these lines of evidence suggest that the true SFR calibration factors cannot be too far from their canonical values. 
\end{abstract}

\keywords{stars: formation -- stars: rotation -- supernovae: general}

\section{Introduction}\label{sec:introduction}

The star formation rate (SFR) is one of the principal parameters characterizing galaxies and their evolution. Much effort has been put into deriving SFR indicators from luminosities spanning wavelengths from the radio to the UV \citep[e.g.,][]{K98,Bell03,Moustakas06,Calzetti07,Salim07,Kennicutt09,Rieke09,Calzetti10}. The efficacy of SFR indicators depends on being able to model the properties of stellar populations at those specific wavelengths. The modeling is typically condensed into ``calibration factors'' that are the ratios of the SFR and the luminosity in a given band, for a given assumed underlying stellar population. For example, the SFR estimated using the H$\alpha$ indicator is the observed H$\alpha$ luminosity, appropriately dust-corrected, multiplied by the H$\alpha$ calibration factor. 

Calculating the calibration factors for the various SFR indicators requires a great deal of complex physics, including challenging problems such as the evolution of massive stars and its dependence on stellar parameters such as mass, metallicity, and rotation; stellar atmospheres; the binary fraction and the resulting interactions; the shape of the stellar initial mass function (IMF); whether star formation is continuous or variable; and so on. The status was summarized by \cite{K98}, who indicated a spread of $\sim 0.1$ dex in H$\alpha$ calibration factors ($\sim 0.3$ dex in the UV) in the literature  when compared for common assumptions of metallicity and IMF\footnote{Reasonable variations of the IMF introduce an overall uncertainty of a factor of $\sim 2$. However, we will test the SFR using supernovae in a way that is weakly affected by changes to the IMF; see Section 4.}. These uncertainties have typically been subdominant compared to the far larger spread between SFR measurements made by different surveys and indicators.  However, the global SFR density and its redshift evolution have been increasingly well mapped in recent years \citep[e.g.,][]{Lilly96,Madau96,Madau98,Steidel99,Hopkins04}. The spread between different SFR density measurements at low redshifts $0<z<1$ is now only several tens of percent \citep[][hereafter HB06]{HB06}\footnote{HB06 also assume a common metallicity and IMF, and notes that IMF variations dominantly affect the accuracy (normalization) instead of the precision (scatter), as long as the IMF variations are universal.}, motivating a renewed look at the calibration uncertainties.

One of the potentially very important physical inputs not considered in many early calibration studies is stellar rotation. Although the standard non-rotating theories of stellar evolution have been highly successful, new data lead to a number of serious discrepancies, and the quantitative effects of rotation have been significantly advanced in the past decade \cite[see, e.g., reviews by][]{Maeder00,Langer12}. Rotation introduces a centrifugal acceleration that reduces the effective gravity, working to reduce the surface temperature of rotating stars. However, often more importantly, rotation also induces internal mixing processes that can have the opposite effect, leading to hotter and more luminous stars \citep[e.g.,][]{Meynet00,Heger00,Brott11}. These, together with rotation induced mass loss, will affect the massive O and B stars that drive the determination of the SFR calibration factors. 

Recently, \cite{E12} presented evolutionary tracks of single stars at solar metallicity, including a complete grid of tracks that have a detailed treatment of axial rotation (the ``ROT'' tracks, to distinguish from the non-rotating ``NOROT'' tracks). In the ROT tracks, they assume initial rotations of $v_{\rm ini} = 0.4 v_{\rm crit}$, where $v_{\rm crit}$ is the velocity where the centrifugal and gravitational accelerations are equal, and the ratio of $0.4$ is motivated by the peak of the velocity distribution observed in young B stars \citep{Huang10}.

If rotation at the level assumed in the ROT tracks of \cite{E12} is ubiquitous among stars, it implies a number of changes and improvements over the NOROT tracks. For example, they provide a better description of the observed properties of Wolf-Rayet stars, including the Wolf-Rayet type ratios, as well as the resulting core-collapse supernova (CCSN) type ratios \citep{Georgy12b}. This echoes studies by \cite{Prantzos03} and \cite{Boissier09} that show the type ratios of core-collapse supernovae (CCSNe) are better matched by rotating stellar models of \cite{Meynet03} and \cite{Maeder04}. The new ROT tracks also show considerably improved agreement with the positions of evolved yellow and red supergiants on the HR diagram, as observed in M~33 \citep{Drout12} and the Large Magellanic Cloud \citep{Neugent12}. Similarly, it has been shown that the increased mass loss rates result in better agreement with the observed yellow supergiant CCSN progenitors \citep{Georgy12}. Finally, using the Starburst99 \citep{Leitherer99} evolutionary synthesis code, \cite{Levesque12} have shown that the ionizing radiation field produced by a population of ROT stars is considerably harder than those produced by a population of NOROT stars. The difference reaches an order of magnitude in the ionizing $\lambda <228 \, {\rm\AA}$ flux. They also note that the bolometric luminosity of the ROT population is higher by a factor $\sim 2$ for the first $\sim 10$ Myr. 

Such changes to the photon output of stellar populations will impact SFR measurements that are derived from observed galaxy luminosities. We quantitatively assess whether these proposed changes are consistent with observations of star formation and CCSN rates.  

We quantify for the first time the effects of stellar rotation as modeled by \cite{E12} on SFR calibration factors. Previously, \cite{Leitherer08} used the rotational stellar tracks of \cite{Meynet00} and reported $\sim 25$\% reductions in SFR calibration factors. We base our calculations on the newer ROT tracks of \cite{E12} and find larger effects. 

We then quantitatively investigate whether the new SFR calibrations are consistent with recent SFR and CCSN data. Since CCSNe mark the ends of the most massive stars formed in a star formation burst, their occurrence is an excellent proxy for recent star formation. CCSNe are particularly suited for our purposes of testing the effects of the ROT tracks on SFR calibrations, because CCSNe provide a benchmark SFR that is minimally affected by stellar rotation: in other words, while stellar rotation is a major parameter affecting what {\it type} of CCSN a collapsing star becomes, the total {\it number} of CCSNe, when all Types II and Ibc are included, likely remains only weakly affected. In addition, the SFR-CCSN comparison is only weakly dependent on the IMF, which allows us to avoid the large uncertainty on the SFR introduced by IMF variations.

We focus on both the cosmic and more local distance regimes. For the latter, we make new calculations of the SFR in the nearby Universe within $11$ Mpc (we henceforth refer this volume simply as the ``local volume''), update the discovery list of local CCSNe, and perform a comparison of the two. The local volume provides a unique test that is complementary to comparisons at cosmic distances, since it is dependent on different inputs with different systematic effects. The completeness of CCSN searches has also improved in recent years, enabling a meaningful analysis with decent statistics. For example, CCSNe in the local volume have been recently utilized in the study of dim CCSNe \citep{Horiuchi11}, the mass range of CCSNe \citep{Botticella12}, and strongly dust-obscured CCSNe \citep{Mattila12}. Here, we use it to study the SFR calibration factors. 

The paper is organized as follows. In Section \ref{sec:tracks}, population synthesis results are presented and the SFR calibration factors are calculated. In Section \ref{sec:implicationSFR}, the effects on SFR measurements are discussed, for both the cosmic and local distances. We then study the comparison with observed CCSN rates in Section \ref{sec:implicationCCSN}. We finish with discussion and a summary in Section \ref{sec:summary}. Throughout, we adopt the standard $\Lambda$CDM cosmology with $\Omega_m=0.3$, $\Omega_\Lambda=0.7$, and $H_0=75 \, {\rm km \, s^{-1} \, Mpc^{-1}}$.

\section{Effects of new stellar tracks on SFR estimators}\label{sec:tracks}

\subsection{Integrated Properties of Stellar Populations}

We use the PEGASE.2 evolutionary synthesis code \citep{FR99} to investigate integrated properties of synthetic stellar populations. The default stellar evolutionary tracks come mainly from the Padova group \citep[][henceforth ``Padova96'']{Alongi93,Bressan93,Fagotto94a,Fagotto94b,Fagotto94c,Girardi96}. These are non-rotational stellar tracks covering a wide range of mass and metallicities. To obtain a complete set of stellar evolutionary tracks, PEGASE.2 combines  asymptotic giant branch (AGB), white dwarf, and unevolving stellar evolutions by various authors \citep{GD93,AB97,Chabrier97}. In addition, the model stellar atmospheres and spectral libraries of \cite{Lejeune97,Lejeune98} are adopted. Throughout, we adopt a fixed metallicity of $Z=0.014$ in order to match the metallicity of the tracks by \cite{E12}. Furthermore, we adopt the supernova model B of \cite{WW95}, a close binary fraction of 0.05, no galactic winds, no treatment of dust extinction, and a minimum CCSN mass of $8 M_\odot$. In PEGASE.2, the supernova model and the minimum mass are only used for calculations of metal enrichment. We checked that changing these parameters do not change the integrated population emission properties noticeably. Unless otherwise stated, we adopt a Salpeter IMF with slope $dN/dM \propto M^{\alpha}$ with $\alpha = -2.35$ over the mass range $0.1$--$100 M_\odot$ \citep{S55}. 

The NOROT and ROT tracks of \cite{E12} are for fixed metallicity ($Z=0.014$) and extend to the core He-flash, early-AGB, and the end of the core C burning for low, intermediate, and massive stars, respectively. Since we are interested in the first $\sim 100$ Myr of evolution and in bands where massive stars dominate, these tracks are generally sufficient. For example, the central evolution of massive stars beyond C burning is usually short enough that the surface properties are no longer modified, and we also do not need to include white dwarf tracks. As was assumed in \cite{Levesque12}, we do not make adjustments to the stellar spectral libraries for rotation, i.e., we use the same libraries for both NOROT and ROT tracks. This is likely not a bad approximation because the libraries are empirically calibrated against observed stellar data. However, we caution that in the absence of a recalibration, the mapping of rotational track parameters to standard stellar libraries may have hidden systematics. Finally, we adopt the same supernova, dust, binary, and IMF assumptions as described above for the Padova96 population. For comparison purposes, we also consider the non-rotating tracks by \cite{S92} and \cite{LS01}, henceforth ``Geneva01''.

\begin{figure}[tb]
\centering\includegraphics[width=\linewidth,clip=true]{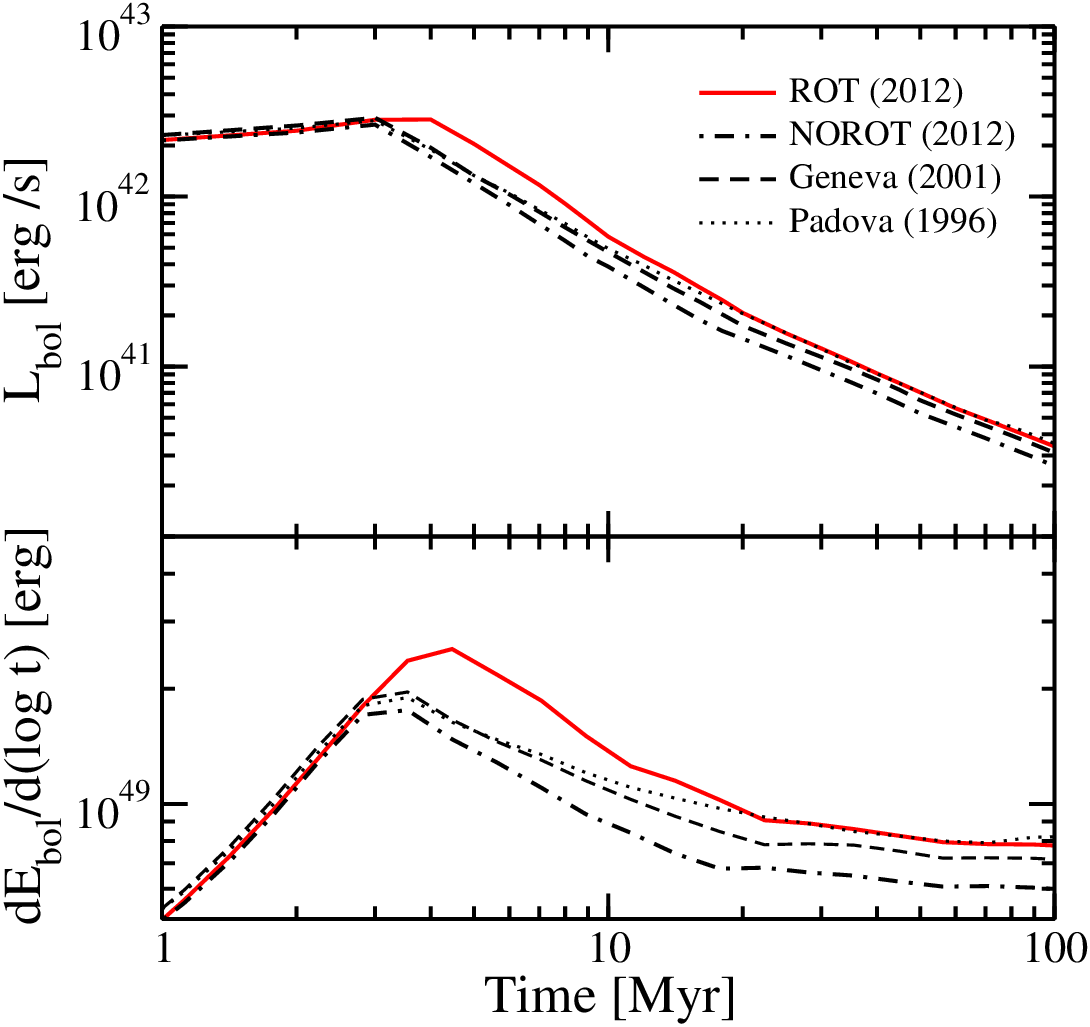}
\caption{\label{fig:burstLbol} Top panel: bolometric luminosity ($dE_{\rm bol}/dt = L_{\rm bol}$) following a burst of star formation with a fixed stellar mass of $10^6 M_\odot$, for the stellar tracks of Padova96 (non-rotating; dotted), Geneva01 (non-rotating; dashed), and for \cite{E12}: NOROT (non-rotating; dot-dashed) and ROT (rotating; red solid). Bottom panel: the luminosity per logarithmic time bin ($dE_{\rm bol}/d{\rm log} \,t $), illustrating the nearly constant energy output after reaching a peak at $\sim 4$ Myr when the most massive stars begin to disappear due to core collapse.}
\end{figure}

We first simulate an instantaneous burst of star formation with a fixed total mass in stars of $10^6 M_\odot$ and follow the evolution of this stellar population up to $100$ Myr. Such a coeval stellar population can be used for studying different stellar samples as the population ages. We use a fixed initial interstellar metallicity and constant stellar metallicity, both of $Z=0.014$, so that the grids of \cite{E12} tracks are not left. Figure \ref{fig:burstLbol} shows the resulting bolometric luminosity. We confirm previous works \citep[e.g.,][]{LE11,Levesque12} and find that the ROT population is $0.4$--$0.5$ mag more luminous than the NOROT population. This difference is a consequence of rotation-induced mixing. This allows mixing of hydrogen into the convective core, allowing stars to burn more hydrogen. The larger core delivers more energy output, and the main-sequence lifetimes are increased by about $25$\%. These echo the results of \cite{Brott11}. We continue to observe this difference out to $100$ Myr, which is consistent with the fact that the rotational effects are observed for all stars above $\sim 2 M_\odot$ \citep{E12}. Here and elsewhere, stellar mass refers to the initial stellar mass at formation, not the current-day mass that is smaller due to mass loss. In the bottom panel, we show how the energy output appears in bins of logarithmic time. 

\begin{figure}[tb]
\centering\includegraphics[width=\linewidth,clip=true]{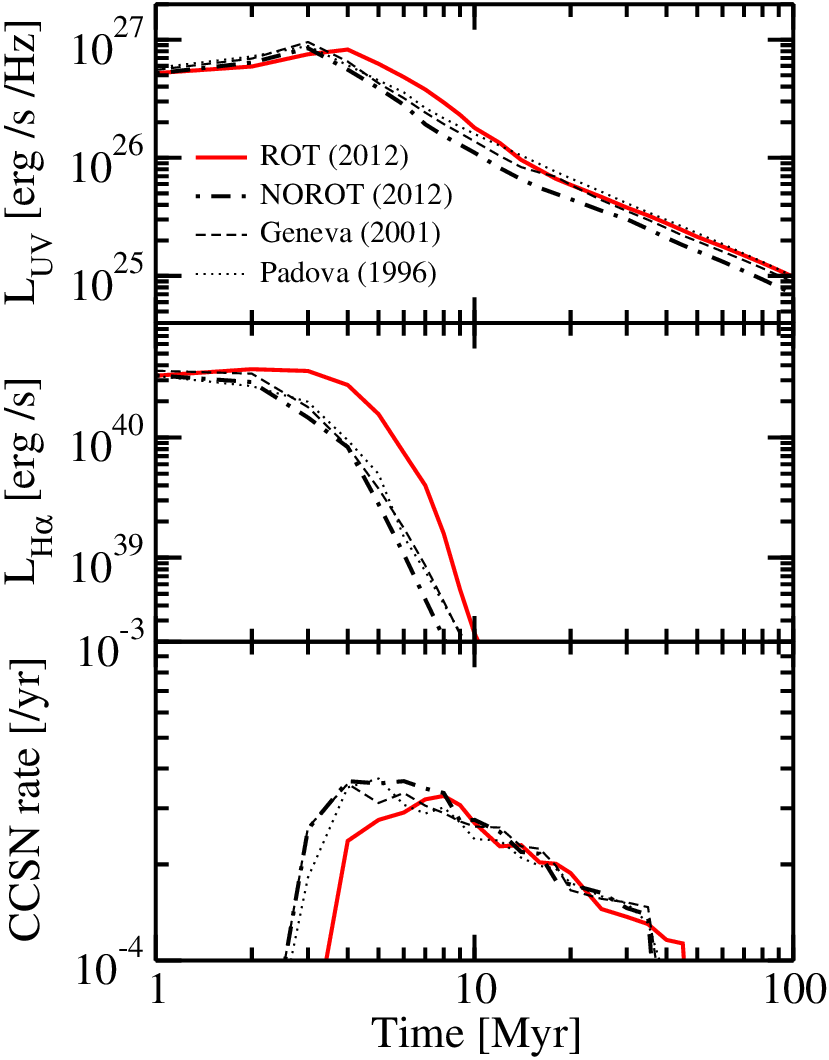}
\caption{\label{fig:burstUVHa} Quantities following a burst of star formation: UV ($0.2{\rm \mu m}$) luminosity (top panel), H$\alpha$ luminosity (middle panel), and the CCSN rate (bottom panel). Star formation burst properties are same as in Figure \ref{fig:burstLbol}. The UV luminosity evolves similarly to the bolometric luminosity, since a wide range of stars ($\sim 3$--$20 M_\odot$) contribute to the UV band. On the other hand, the H$\alpha$ luminosity is powered by the most massive stars ($\gtrsim 20 M_\odot$) and its behavior is qualitatively different.}
\end{figure}	

The NOROT population evolves similarly to those based on the older non-rotating tracks of Padova96 and Geneva01, but after several Myr becomes slightly underluminous. This is due to the updated overshooting and mass-loss prescriptions of the NOROT tracks. They decrease the main sequence lifetime of $M<30 M_\odot$ mass stars and increases the main sequence lifetime of $M> 30 M_\odot$ stars, respectively \citep[see Figure 6 of][]{E12}. Therefore, after a few Myr when the most massive stars have disappeared, the difference between populations are dominated by $M < 30 M_\odot$ stars, and the NOROT population is expected to become dimmer compared to the Padova96 and Geneva01 populations. However, the differences are small, and we will henceforth quote effects of rotation relative to the updated non-rotational NOROT tracks unless specifically stated otherwise. 

In Figure \ref{fig:burstUVHa}, we show the UV ($0.2{\rm \mu m}$) luminosity, the H$\alpha$ luminosity, and the CCSN rate following the same star formation burst. The UV luminosity evolution is similar to that of the bolometric luminosity, because the dominant contribution comes from a wide mass range of stars; for our adopted tracks, $\sim 3$--$20 M_\odot$ stars. On the other hand, the H$\alpha$ luminosity shows a qualitatively different behavior. This is because the H$\alpha$ emission is largely sensitive to photons shortward of the Lyman limit that dominantly arise from massive young stars with $M \gtrsim 20 M_\odot$. The H$\alpha$ thus falls when these massive stars disappear in core collapse. The progenitors of CCSN are assumed to be massive $M>8 M_\odot$ stars. These stars fall in between the mass range of stars that dominantly power the UV and H$\alpha$ fluxes. Thus, the CCSN rate rises when the most massive $100 M_\odot$ stars core collapse and falls when the $8 M_\odot$ stars core collapse. 

For the CCSN rate, the difference between tracks is only due to the different main sequence lifetimes of the CCSN progenitors. This is because we assume all stars with $M>8 M_\odot$ core collapse to become CCSNe, for both non-rotating and rotating stars. As a result, the ROT tracks with their longer main-sequence lifetimes show a later rise in the CCSN rate. On the other hand, the UV and H$\alpha$ properties also strongly reflect the differences in luminosity and surface temperature. 

\subsection{Star Formation Rate Calibration Factors}

\begin{figure}[tb]
\centering\includegraphics[width=\linewidth,clip=true]{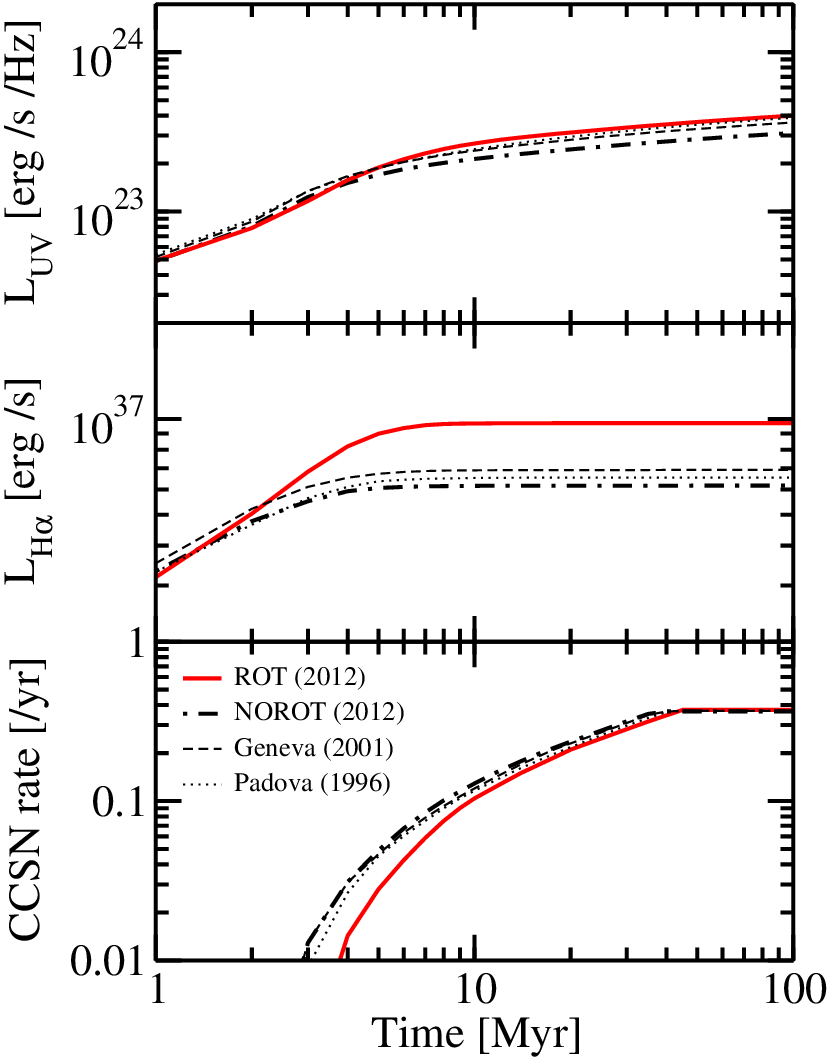}
\caption{\label{fig:contUVHa} Same as Figure \ref{fig:burstUVHa} but for a continuous star formation rate of $50 \, {\rm M_\odot \, yr^{-1}}$. The H$\alpha$ reaches equilibrium earliest since they are sourced by the most massive stars only. The progenitors of CCSNe are $M>8M_\odot$ stars that reach equilibrium later, while the UV has some contribution from stars with lifetimes longer than $100$ Myr. }
\end{figure}	

A variety of indicators are used as tracers of recent star formation activity, all directly or indirectly probing the current population of massive stars \citep[e.g.,][]{K98,Bell03,Moustakas06,Calzetti07,Calzetti10,Salim07,Kennicutt09,Rieke09}. The UV stellar continuum and the H$\alpha$ nebular emission have traditionally been extensively used due to their accessibility. We compute the UV and H$\alpha$ calibration factors using PEGASE.2 assuming standard assumptions of constant SFR for at least $100$ Myr. This is the continuous star formation approximation, which provides enough time to allow the birth and death of massive stars that dominate the luminosity in the UV and H$\alpha$ to equilibrate. It is applicable at least at low-$z$ where the fraction of the total SFR contributed from young starbursts is thought to be $<10$\% \citep[e.g.,][]{Salim05}. 

In Figure \ref{fig:contUVHa}, we show the UV ($0.2{\rm \mu m}$) luminosity, the H$\alpha$ luminosity, and the CCSN rate, all for a continuous SFR of $50 \, {\rm M_\odot \, yr^{-1}}$. As can be seen, the H$\alpha$ luminosity and CCSN rate comfortably reach equilibrium before $100$ Myr. By contrast, the UV is still marginally increasing because the UV has contributions from stars with main-sequence lifetimes longer than $100$ Myr. However, this is not a large fraction. The UV luminosity only increases by another $\sim 15$\% over the next few hundred Myrs before reaching equilibrium. It should be mentioned that the CCSN rate for different stellar tracks all tend to the same value. This is because we have defined the CCSN rate by the mass range of stars undergoing core collapse (see Section \ref{sec:implicationCCSN} for more details).

\begin{figure}[tb]
\centering\includegraphics[width=0.97\linewidth,clip=true]{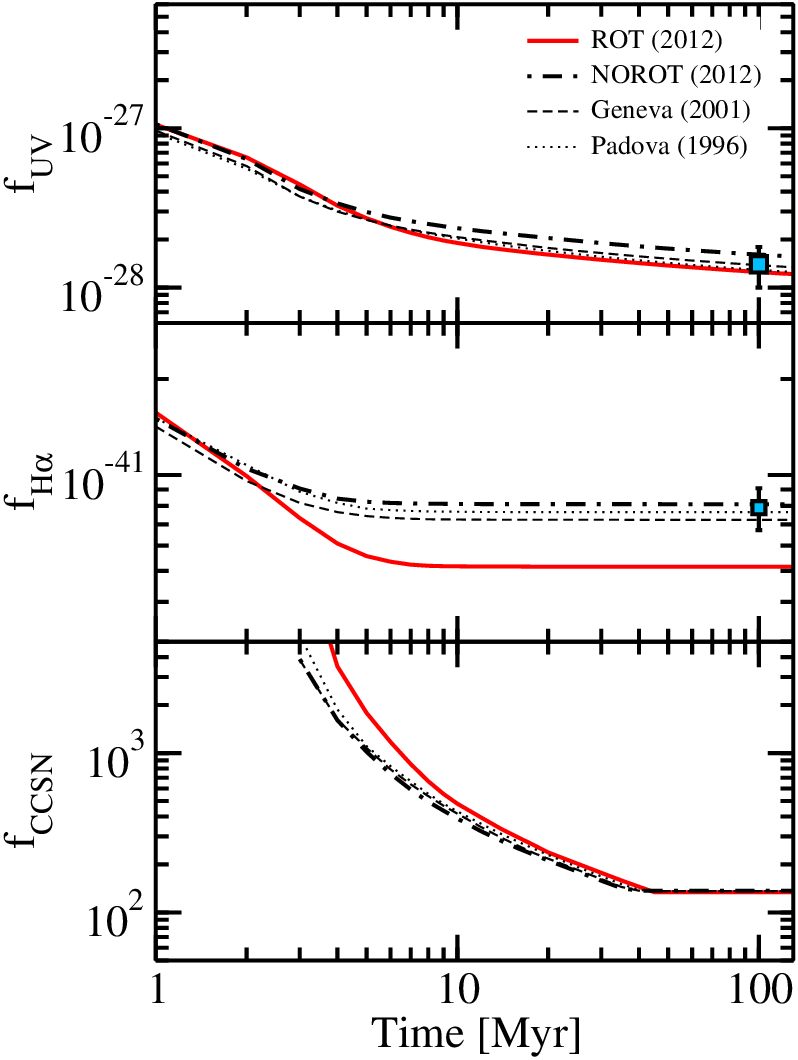}
\caption{\label{fig:calibrations} UV (top panel) and H$\alpha$ (middle panel) calibration factors calculated under the continuous star formation approximation starting at time $t=0$. Calibration factors are defined such that ${\rm SFR} = f_x L_x$, i.e., in units of $(M_\odot {\rm yr}^{-1} ) / ({\rm erg \, s^{-1} \, Hz^{-1}} )$ and $(M_\odot \, {\rm yr}^{-1}) / ({\rm erg \, s^{-1}} ) $ for the UV and H$\alpha$, respectively.  For comparison, we mark the values quoted in \cite{K98} as a data point. In the bottom panel we show the CCSN calibration factor (see Section \ref{sec:implicationCCSN} for details).}
\end{figure}	

We define the calibration factor $f_x$ by
\begin{equation}
{\rm SFR} = f_x L_x,
\end{equation}
where SFR is in units $M_\odot \, {\rm yr}^{-1}$, and $L_x$ is the dust-corrected luminosity in units $\rm erg \, s^{-1}$ for the H$\alpha$ and $\rm erg \, s^{-1} \, Hz^{-1}$ for the UV. Results for the Padova96, Geneva01, NOROT, and ROT tracks are shown in Figure \ref{fig:calibrations} where their time evolutions are shown, and their values for $100$ Myr after the onset of star formation are summarized in Table \ref{table:calibration}. First, we see that the new NOROT calibrations are similar to the Padova96, Geneva01, and \cite{K98} calibrations, although they are slightly larger. For example, compared to the commonly used values of \cite{K98}, the UV calibration is larger by $\sim 10$\%. Second, the ROT tracks yield smaller calibrations. For example, the ROT UV and H$\alpha$ calibrations are smaller by $30$\% and $40$\% when compared to the NOROT calibrations. These follow from the increased main-sequence lifetimes and luminosities of rotating stars. The widely adopted calibrations of \cite{K98} are shown for comparison with error bars that reflect the full range of values reported by \cite{K98}. Most of our calibrations from Table \ref{table:calibration} are within this range, but the ROT H$\alpha$ calibration is significantly lower. 

We change the PEGASE.2 input parameters to test the robustness of these results. One of the most important inputs for calculating the calibration factor is the IMF. Generally, a shallower (steeper) IMF results in relatively more (less) massive stars and smaller (larger) calibration factors. We consider three IMFs for comparison: the ``SalA'' IMF, as proposed by \cite{BG03}, with the Salpeter shape above $0.5 M_\odot$ (high-mass gradient of $-2.35$) and a suppression below $0.5 M_\odot$ (low-mass gradient of $-1.5$); the ``BG'' IMF, also by \cite{BG03}, with a high-mass gradient of $-2.15$ and a low-mass gradient of $-1.5$; and finally, the IMF of \cite{Kroupa01}. The SalA, BG, and Kroupa IMFs respectively yield UV calibration factors that are $\sim 0.77$, $\sim 0.50$, and $\sim 0.83$ times those of the Salpeter IMF, all for the NOROT tracks. The H$\alpha$ changes slightly more: by $\sim 0.77$, $\sim 0.42$, and $\sim 0.79$, respectively. 

\begin{deluxetable}{lcc}
\tablecaption{Summary of $Z=0.014$ Calibration Factors for Salpeter IMF at $100$ Myr (continuous SFR) \label{table:calibration}}
\tablehead{ 		&  $f_{\rm UV}$ 		& $f_{\rm H\alpha}$		 	}
\startdata
Kennicutt (1998)	& $1.4 \times 10^{-28}$	& $7.9 \times 10^{-42}$	\\ 
\tableline
Padova (1996)		& $1.3 \times 10^{-28}$	& $7.7 \times 10^{-42}$	\bigstrut[t]	\\
Geneva (2001)		& $1.4 \times 10^{-28}$	& $7.2 \times 10^{-42}$	\\
NOROT (2012)		& $1.6 \times 10^{-28}$	& $8.1 \times 10^{-42}$	\\
ROT (2012)		& $1.2 \times 10^{-28}$	& $5.2 \times 10^{-42}$
\enddata
\tablecomments{Calibration factors defined as $f_x = {\rm SFR} / L_x$. The H$\alpha$ calibration has units of $(M_\odot \, {\rm yr}^{-1}) / ({\rm erg \, s^{-1}} ) $ and the UV calibration has units of $(M_\odot {\rm yr}^{-1} ) / ({\rm erg \, s^{-1} \, Hz^{-1}} )$. }
\end{deluxetable}

Repeating the exercise with the ROT tracks, very similar changes are observed. For example, the SalA, BG, and Kroupa IMFs yield UV calibration factors that are $\sim 0.78$, $\sim 0.49$, and $\sim 0.83$ times those of the Salpeter IMF, and H$\alpha$ calibration factors that are $\sim 0.77$, $\sim 0.41$, and $\sim 0.79$ times those of the Salpeter IMF. Therefore, while the absolute values of the calibration factors change as a result of the IMF shape, the relative importance of stellar rotation is always a reduction of $30$\%--$40$\%. We also stress here that the IMF does not strongly impact the comparison with the CCSN rate (see Section \ref{sec:implicationCCSN}).

Changing the mass range of the IMF also makes a difference to the SFR calibration factors. For example, increasing the high-mass cutoff from our canonical $100 M_\odot$ to $120 M_\odot$ leads to more massive stars (for a fixed total stellar mass) and a $3$\% and $15$\% decrease in the UV and H$\alpha$ NOROT calibrations, respectively. Some authors adopt an upper limit of $60 M_\odot$ \citep[e.g.,][]{Kobayashi12} which would have the opposite effect: $12$\% and $60$\% increases, respectively. We note here that the use of $60 M_\odot$ necessarily underestimates the H$\alpha$ calibration because the important massive stars are not included. Once again, the effect of rotation is a $30$\%--$40$\% reduction in the calibration factors regardless of the mass range. This is partly due to the assumption of a rotation velocity $v_{\rm ini} = 0.4 v_{\rm crit}$ for all stellar masses of the ROT tracks.

Metallicity is also important since the evolution of massive stars are strongly affected by metal-driven mass loss \citep[e.g.,][]{Kudritzki89}. Generally, low-metallicity stars lose less mass, are more compact, and more luminous, leading to smaller SFR calibration factors. For the Padova96 tracks and a Salpeter IMF, changing the metallicity to $Z=0.001$ ($Z=0.1$) causes the UV calibration to change by $-2$\% ($+10$\%) when compared to our canonical $Z=0.014$ estimates, and the H$\alpha$ calibration to change by $-30$\% ($+50$\%). We do not quantitatively study rotation effects for different metallicities since the \cite{E12} tracks are only for a single metallicity. However, rotational mixing is less efficient in high-metallicity stars, and high-metallicity stars also lose more mass through metal-driven mass loss which slows the star down \citep{Brott11}. At the other extreme, low-metallicity massive stars are so efficiently mixed that they may evolve chemically homogeneously \citep{Brott11}. Other parameters such as the minimum mass for CCSNe, the supernova chemical abundance model, and the presence of Galactic winds, only slowly affect the luminous outputs of stars through feedback of chemically enriched gas. 

Other star formation indicators that are routinely used include the [OII]$\lambda 3727$ forbidden line, infrared, and radio emissions. We do not calculate calibrations for these directly. However, because they are often associated with either the UV or H$\alpha$, we can discuss how they are affected by stellar rotation. For example, the [OII] luminosity is not directly coupled to the ionizing luminosity, but as a SFR tracer it is empirically calibrated based on H$\alpha$, often using samples of nearby galaxies \citep{K92,Moustakas06}. This is non-trivial since the [OII] to H$\alpha$ ratio depends on the luminosity, metallicity, and also obscuration. But it implies that changes to the H$\alpha$ calibration factor will be carried through to a change in the [OII] calibration factor. 

The infrared luminosity is an indirect tracer of star formation arising from UV continuum that is absorbed and re-radiated by dust. It is complementary to the UV and H$\alpha$ that are absorbed by dust, and calibrations have been investigated by many authors \citep{Sauvage92,Bell03,Calzetti07,Calzetti10,Rieke09}. It requires sufficient optical depth of the dust in the star formation region, such that the $L_{\rm IR}$ provides a calorimetric measure of the star formation. However, in practice, the infrared is contaminated by dust heated by old stellar populations. For spiral galaxies the contamination to the far-infrared ($1$--$1000 \mu$m) can be as high as $70$\% \citep[e.g.,][]{LP87}, and modern calibrations therefore utilize shorter wavelengths where contamination is less of an issue \citep{Calzetti10}. So, rotation effects on the low-mass stellar population must also be considered to some degree, depending on the galaxy sample and exact wavelength used. According to \cite{E12}, rotation affects low mass stars ($0.8$--$2 M_\odot$) less than the massive stars, which means that the infrared calibrations will be less affected by stellar rotation than the UV or H$\alpha$.

Radio continuum emission is another indirect tracer of star formation. It is insensitive to dust attenuation, but it relies on the complex and poorly understood physics of cosmic ray production and confinement in supernova remnants. The $1.4$ GHz SFR calibration of \cite{Bell03} partly avoids this difficulty by empirically using the observed tight radio-FIR correlation, and calibrating to be consistent with the FIR calibration of \cite{K98}. This calibration is a factor $\sim 2$ smaller than the calibration of \cite{Condon92} that is calculated from first principles. Rotation will reduce the radio calibrations of \cite{Bell03} through the infrared calibration, but will not affect those of \cite{Condon92}.

To conclude, recent modeling of stellar rotation imply reductions in SFR calibration factors by $30$\% for the UV and $40$\% for the H$\alpha$, where we have compared the ROT population to the NOROT population. Comparing the ROT calibrations to the widely used non-rotating calibrations of \cite{K98}, the reductions are $10$\% and $40$\% for the UV and H$\alpha$, respectively. 

\section{Implications for SFR estimates} \label{sec:implicationSFR}

The reductions to the SFR calibration factors due to stellar rotation are somewhat smaller than the reductions caused by variations in the IMF ($\sim 0.77$ for SalA and $\sim 0.50$ for BG). Nevertheless, they have testable consequences. As we discuss in Section \ref{sec:implicationCCSN}, observations of CCSNe provide a benchmark SFR whose comparison with the directly measured SFR is minimally dependent on the IMF. Also, because stellar rotation affects the H$\alpha$ indicator more than the UV, we can use the UV/H$\alpha$ ratio as a test.

\subsection{SFR at Cosmic Distances}

At cosmic distances, the SFR has been measured by many groups using a variety of indicators. Comparisons of data are complicated by different galaxy selection biases, different assumptions of dust attenuation, different luminosity cuts, cosmic variance, and other issues \cite[e.g.,][]{Hopkins01,Hopkins04}. To address these, \cite{HB06} selected recent SFR data, converted them to a common cosmology, dust correction scheme, IMF, and calibration assumptions. Using their compilation, they found a best-fit cosmic SFR density with a $z=0$ normalization of $2.1^{+0.4}_{-0.3} \times 10^{-2}\, {\rm M_\odot \, Mpc^{-3} \, yr^{-1}}$ ($1 \sigma$ errors) for a Salpeter IMF and SFR calibrations of \cite{K98}. 

The best-fit uncertainty of \cite{HB06} is $\pm 20$\%--$30$\% in the most constrained redshift range of $0 < z < 1$. In this redshift range, the HB06 fit is based mostly on UV-derived SFR measurements, except for one H$\alpha$ measurement at $z \approx 0.01$ by \cite{Hanish06}. Changing the UV calibration from their adopted \cite{K98} values to our ROT values only results in a mild $\sim 10$\% reduction in the SFR, which is smaller than the fit uncertainty. At larger redshifts, a wider range of indicators are used in the HB06 fit. Since stellar rotation affects different indicators by different degrees, stellar rotation will contribute additional spread as well as reduce the SFR normalization. However, the HB06 best-fit uncertainty is factors of $\sim 2$ or more at these larger redshifts, so rotation would not be a dominant contributor to the uncertainty.

\subsection{SFR at Local Distances}
In contrast to the cosmic SFR, studies of the local volume within $\sim 10$ Mpc have the advantage that all use almost identical galaxy samples. The $11$ Mpc H$\alpha$ and Ultraviolet Galaxy Survey ($11$HUGS) is designed to provide a census of the SFR in the volume within $11$ Mpc using the H$\alpha$ indicator \citep{K08}. The distance limit was set to have both statistically significant galaxy numbers ($436$ galaxies) and galaxy completeness ($M_B = -15$ mag at the distance limit $11$ Mpc; however, the number of galaxies as a function of distance does not rise as fast as expected from a constant density of galaxies, as shown in Figure~\ref{fig:localSN}). These local galaxies have also been observed by the $GALEX$ and $Spitzer$ satellites, providing multi-band UV and IR photometry: these are the GALEX Nearby Galaxy Survey \citep[$390$ galaxies, including $363$ from the $11$HUGS;][]{G07,Lee11} and the composite Local Volume Legacy (LVL) survey \citep[$258$ galaxies, including $231$ from the $11$HUGS;][]{Dale09}. 

Using a sample of $\sim 300$ local galaxies, \cite{Lee09} carefully investigated the H$\alpha$- and UV-SFR. After correcting for dust, they find that the H$\alpha$-SFR tend to be smaller than the UV-SFR, log$({\rm SFR_{UV}} / {\rm SFR_{H\alpha}})  \approx 0.13$, i.e., a factor of $1.35$, for the most star-forming galaxies with H$\alpha$-SFR $> 10^{-1.5} \, {\rm M_\odot yr^{-1}}$. In another study, \cite{Botticella12} studied $312$ local galaxies and derived dust-corrected H$\alpha$ and UV luminosities of $L_{\rm H\alpha} = (99 \pm 5) \times 10^{41} \, {\rm erg \, s^{-1}}$ and $L_{\rm UV} =(88 \pm 6) \times 10^{28} \, {\rm erg \, s^{-1} \, Hz^{-1}}$, respectively. Using the calibrations of \cite{K98}, they find H$\alpha$- and UV-SFR values of $78 \pm 4 \, {\rm M_\odot yr^{-1}}$ and $123 \pm 8 \, {\rm M_\odot yr^{-1}}$, respectively.

If we apply the ROT calibration factors to the local luminosities of \cite{Botticella12}, the difference between the H$\alpha$- and UV-SFR will widen from their nominal ratio of ${\rm UV}$-${\rm SFR}/{\rm H\alpha}$-${\rm SFR} \approx 1.6$ to ${\rm UV}$-${\rm SFR}/{\rm H\alpha}$-${\rm SFR} > 2$. This difference is considerably larger than the reported measurement errors, and disfavors ubiquitous stellar rotation in the local galaxies. However, it is premature to conclude that stellar rotation is not allowed by local SFR estimates. On an individual galaxy basis, different indicators give discrepant SFR estimates, sometimes up to factors of $\sim 10$ \citep{Hopkins03}. Although the difference is much reduced for larger samples of galaxies, systematic uncertainties in flux measurements, dust attenuation corrections, stochasticity in the formation of high mass stars, variations in the IMF, and departures from standard recombination calculations can still result in systematic offsets between the two indicators \citep[e.g.,][]{Buat87,Buat92,Hopkins01,Buat02,Bell01,I04,Salim07,Bothwell09}. Of these, the most plausible source of problems is dust corrections. Indeed, \cite{Botticella12} argue the UV-H$\alpha$ discrepancy arises due to uncertain dust corrections of atypical galaxies that dominate the SFR budget; for example, high inclination galaxies (e.g., M~$82$) or galaxies with large foreground Galactic extinction (e.g., NGC~$6946$). 

\subsection{Total SFR within $11$ Mpc}

We derive the total SFR within the $11$ Mpc volume and compare to the cosmic SFR density. For this we must assume a local overdensity factor \citep[e.g.,][]{Bothwell11}. For example, at distances of $1$--$5$ Mpc, \cite{Drozdovsky08} demonstrate more than factor two overdensity over the cosmic SFR. \cite{K04} studied the $B$-band luminosity density of the local $8$ Mpc and found it to be $1.7$--$2.0$ times the global luminosity density as measured by the Sloan Digital Sky Survey and the Millennium Galaxy Catalogue. Using an overdensity factor of $1.7$, and adjusting to the NOROT calibration, the $z=0$ extrapolation of the HB06 best-fit cosmic SFR density predicts a total SFR within $11$ Mpc of $230^{+50}_{-40} \, {\rm M_\odot yr^{-1}}$. Similarly, the ROT calibration yields $180^{+40}_{-30} \, {\rm M_\odot yr^{-1}}$. In both estimates we use the UV calibration, since the HB06 compilation in the redshift range $0 < z < 1$ is dominantly based on the UV indicator. 

We directly estimate the total SFR within $11$ Mpc by starting with the $11$HUGS + $GALEX$ galaxy catalog and applying the same two-tier Galactic latitude cut as is used in the LVL survey, namely, $|b| \geq 20^\circ$ for distances $<3.5$ Mpc and $|b| \geq 30^\circ$ for distances of $3.5$ to $11$ Mpc. This leaves a sample of $282$ galaxies with positive H$\alpha$ and FUV flux measurements. The fraction of the $11$ Mpc sphere included after these cuts is $\sim 51$\%, and we estimate the total SFR by summing the SFR of the galaxies and dividing by the volume fraction.

{{\renewcommand{\arraystretch}{1.1}
\begin{deluxetable}{lr}
\tablecaption{Summary of Total SFR Estimates within $11$ Mpc \label{table:results}}
\tablehead{ 						&SFR ($M_\odot$ yr$^{-1}$)	}
\startdata
Cosmic extrapolation (NOROT)		& $230 \pm 50$	\\	
Direct H$\alpha$ (NOROT)			& $120 \pm 20$	\\
Direct UV (NOROT)					& $170 \pm 20$	\\
\tableline	
Cosmic extrapolation (ROT)			& $180 \pm 40$	\\
Direct H$\alpha$ (ROT)				&  $ 80 \pm 10$	\\
Direct UV (ROT)					& $130 \pm 20$	\\
\tableline	
\tableline	
CCSN with 08S-like objects			& $360^{+120}_{-90}$	\bigstrut[t] 		\\
CCSN without 08S-like objects			& $270^{+110}_{-80}$
\enddata
\tablecomments{Comparison of the total SFR within $11$ Mpc, i.e., sky-coverage corrected where appropriate, derived from cosmic SFR, local UV, local H$\alpha$, and local CCSN measurements. For estimates from the cosmic SFR (cosmic extrapolation), a local overdensity of a factor $1.7$ is assumed. For estimates from the local UV measurements (Direct UV), Equations (\ref{eq:GALEXcalibration}) and (\ref{eq:GALEXcalibrationROT}) are adopted for the non-rotating (NOROT) and rotating (ROT) estimates, respectively. For estimates from the local H$\alpha$ measurements (Direct H$\alpha$), calibration factors of Table \ref{table:calibration} are adopted. For estimates from local CCSNe, Equation (\ref{eq:CCSNrate}) with a mass range of $8$--$40 \, M_\odot$ is adopted; the upper mass limit is not important (see Section \ref{sec:implicationCCSN}). 08S-like objects refer to optical transients that may or may not be true core collapses (see Section \ref{sec:localCCSN} for discussions).}
\end{deluxetable}
}

The H$\alpha$ luminosities are corrected for the [NII] line contamination, underlying stellar absorption, and Galactic foreground extinction following \cite{K08}. We correct for internal attenuation using the empirical scaling correction with the host galaxy $B$-band magnitude \citep{Lee09} of
\begin{equation}
A_{H\alpha}=\left\{ \begin{array}{lll}
0.10 & M_B > -14.5\\
1.971+0.323M_B+0.0134M_B^2 & M_B \leq-14.5\\
\end{array}
\right.
\end{equation}
For the FUV measurements, we correct for internal attenuation based on the tight correlation with the TIR/FUV ratio \citep{Burgarella05},
\begin{equation}
A({\rm FUV}) = -0.028x^3 + 0.392 x^2 + 1.094 x + 0.546,
\end{equation}
applicable for galaxies with recent star formation such as the spirals and irregulars. Here, $x = {\rm log}[ L({\rm TIR}) / L({\rm FUV})_{\rm obs} ] $. When the TIR data give a negative correction, or when TIR data are not available, we use the empirical relation $A({\rm FUV}) = 1.8A({\rm H\alpha})$ of \cite{Lee09}. These corrections are similar to those applied in previous studies \citep{Lee09,Botticella12}. Finally, we modify the UV calibrations of Table \ref{table:calibration} to reflect the central wavelength of the $GALEX$ FUV filter, $0.1532 {\rm \mu m}$. For the NOROT, we find that
\begin{equation} \label{eq:GALEXcalibration}
{\rm log \, SFR\,[M_\odot yr^{-1}] } = {\rm log ( L_{\rm FUV,corr} } {\rm [L_\odot] } ) - 9.51,
\end{equation}
which is very similar to the $GALEX$ calibration reported in \cite{I06} who use the Starburst99 synthesis code \citep{Leitherer99} with the solar metallicity Geneva non-rotating tracks and a Salpeter IMF from $0.1$--$100 M_\odot$. For reference, the equivalent for the ROT is
\begin{equation} \label{eq:GALEXcalibrationROT}
{\rm log \, SFR\,[M_\odot yr^{-1}] } = {\rm log ( L_{\rm FUV,corr} } {\rm [L_\odot] } ) - 9.63.
\end{equation}

We summarize our results in Table \ref{table:results}. The direct SFR estimates are generally smaller than the cosmic extrapolations, but the UV-SFR is consistent within the uncertainties. The H$\alpha$ estimates, on the other hand, fall short of the cosmic extrapolation. One possibility is that we have slightly overestimated the local overdensity, since we adopt an overdensity derived from $\sim 8$ Mpc scales for the entire $11$ Mpc volume.

As mentioned in the previous section, the stochastic nature of star formation can be an issue. Since PEGASE.2 is a deterministic synthesis code, it does not cover the stochastic behavior at low SFR rates. This occurs when the number of massive stars falls below $\sim 10$. However, we are summing over a sufficient number of galaxies with a sufficiently large total SFR that the stochasticity is not a problem. More quantitatively, the H$\alpha$ indicator is related to stars with $M \gtrsim 20 \, M_\odot$. Adopting the Salpeter IMF, $5 \times 10^3 \, M_\odot$ of stellar material must be made for there to be $10$ such stars, and this must continue at least over the lifetime of the massive stars, some $3$ Myr. Thus, the required SFR is at least $\sim 2 \times 10^{-3} \, {\rm M_\odot \, yr^{-1}}$ per galaxy. If we adopt more modern IMFs where the number of low-mass stars are suppressed relative the Salpeter IMF, there will be more massive stars for a given total stellar mass and the required SFR will only decrease.

\section{Comparison to core-collapse supernova rates} \label{sec:implicationCCSN}

The most massive stars in a star formation burst evolve the fastest and then explode as CCSNe. Therefore, observations of CCSNe are excellent proxies of recent star formation that is far less sensitive to modest dust. Different types of CCSNe probe the SFR on different time scales; treated together, CCSNe probe the SFR on time scales similar to the UV indicator. A well-measured CCSN rate yields an estimate of the SFR normalization, and even the observation of a few CCSNe yields lower limits. We use both the cosmic CCSNe and CCSNe in the local volume and discuss how compatible they are with the rotation calibrations.

\begin{deluxetable*}{lllclrrlll}
\tablecaption{Local CCSNe (in Bold, Total 17) and Possible CCSNe (in Italic, Total 5) within $11$ Mpc During 2000 to 2011 Inclusive \label{table:local}} 
\tablehead{ 
SN			& Galaxy			& Type	& $D$ (Mpc)	& R.A.	& \multicolumn{1}{l}{Decl.}	& $b$ (deg)	&\multicolumn{2}{c}{11HUGS}	& LVL	\\
			& 				& 		& 			& 		& 	&	& Full		& LVLcut		& 	 } 
\startdata
{\bf SN~2002ap}	& NGC 0628		& IcPec			& 7.3		& $01\; 36\; 41.7 $	& $15\; 46\; 59$	& $-45.7$ & Y & Y  & Y \\ 
{\it SN~2002bu}	& NGC 4242		& SN?$^{1}$		& 7.4		& $12\; 17\; 30.1$ 	& $45\; 37\; 08$	& $70.3$  & Y & Y  & Y \\  
{\bf SN~2002hh}	& NGC 6946		& IIP   			& 5.9		& $20\; 34\; 52.3$ 	& $60\; 09\; 14$	& $11.7$  & Y & -   & -  \\   
{\bf SN~2003gd}	& NGC 0628		& IIP      			& 7.3		& $01\; 36\; 41.7$	& $15\; 46\; 59$	& $-45.7$  & Y & Y  & Y \\  
{\bf SN~2004am}	& NGC 3034 (M 82)	& IIP 				& 3.5		& $09\; 55\; 52.2$	& $69\; 40\; 47$	& $40.6$  & Y & Y  & Y \\ 
{\bf SN~2004dj}	& NGC 2403		& IIP    			& 3.2		& $07\; 36\; 51.4$	& $65\; 36\; 09$	& $29.2$  & Y & Y  & Y \\ 
{\bf SN~2004et	}	& NGC 6946		& IIP				& 5.9		& $20\; 34\; 52.3$	& $60\; 09\; 14$	& $11.7$  & Y & -   & -  \\
{\bf SN~2005af	}	& NGC 4945		& IIP				& 3.6		& $13\; 05\; 27.5$	& $-49\; 28\; 06$	& $13.3$  & Y & -   & -  \\
{\bf SN~2005at	}	& NGC 6744		& Ic				& 9.4		& $19\; 09\; 46.2$	& $-63\; 51\; 25$	& $-26.1$  & Y & -   & -  \\
{\bf SN~2005cs	}	& NGC 5194 (M 51)	& IIP      			& 8.0		& $13\; 29\; 52.7$	& $47\; 11\; 43$	& $68.6$  & Y & Y  & Y \\
{\bf SN~2007gr	}	& NGC 1058		& Ic  	 			& 9.2		& $02\; 43\; 29.9$	& $37\; 20\; 27$	& $-20.4$  & Y & -   & -  \\
{\it SN~2008S}		& NGC 6946		& SN?$^{2}$		& 5.9		& $20\; 34\; 52.3$ 	& $60\; 09\; 14$	& $11.7$  & Y & -  & -  \\
{\bf SN~2008ax}	& NGC 4490		& IIb				& 8.0		& $12\; 30\; 36.1$	& $41\; 38\; 34$	& $74.9$  & Y & Y  & Y \\
{\bf SN~2008bk}	& NGC 7793		& IIP  			& 3.9		& $23\; 57\; 49.7$	& $-32\; 35\; 30$	& $-77.2$  & Y & Y  & Y \\
{\it NGC300-OT}	& NGC 0300		& SN?$^{3}$		& 2.0		& $00\; 54\; 53.5$ 	& $-37\; 41\; 00$	& $-79.4$  & Y & Y  & -  \\
{\bf SN~2008iz }	& NGC 3034 (M 82)	& II   				& 3.5		& $09\; 55\; 52.2$ 	& $69\; 40\; 47$	& $40.6$  & Y & Y  & Y \\
{\bf SN~2008jb}	& ESO 302-14		& IIP				& 9.6		& $03\; 51\; 40.9$	& $-38\; 27\; 08$	& $-50.9$  & Y & Y  & -  \\
{\bf SN~2009hd}	& NGC 3627 (M 66)	& IIP				& 10.1	& $11\; 20\; 15.0$ 	& $12\; 59\; 30$	& $64.4$  & Y & Y  & Y \\
{\it SN~2010da}	& NGC 0300		& SN?$^{4}$		& 2.0		& $00\; 54\; 53.5$ 	& $-37\; 41\; 00$	& $-79.4$  & Y & Y  & -  \\
{\bf SN~2011dh}	& NGC 5194 (M 51)	& IIb				& 8.0		& $13\; 29\; 52.7$ 	& $47\; 11\; 43$	& $68.6$  & Y & Y  & Y \\
{\it PSN J12304185+4137498} & NGC 4490 & SN?$^{5}$	& 8.0		& $12\; 30\; 36.1$	& $41\; 38\; 34$	& $74.9$  & Y & Y  & Y \\
{\bf SN~2011ja}	& NGC 4945		& IIP				& 3.6		& $13\; 05\; 27.5$ 	& $-49\; 28\; 06$	& $13.3$  & Y  & -    & - 
\enddata
\tablecomments{$^{1}$\cite{Thompson09,Smith11,Szczygiel12b}, $^{2}$\cite{Prieto08,Prieto09,Smith09,Botticella09,Pumo09,Prieto10,Szczygiel12a}, $^{3}$\cite{Bond09,Berger09,Prieto09,Kashi10,Kochanek12}, $^{4}$\cite{Elias10}, $^{5}$\cite{Cortini11,Fraser11,Magill11}}
\end{deluxetable*}

The CCSN rate, $\dot{N}_{\rm CCSN}$, is related to the SFR, $ \dot{M}_*$, by $\dot{N}_{\rm CCSN} = \dot{M}_*/f_{\rm CCSN}$, where $f_{\rm CCSN}$ is calculated as
\begin{equation}\label{eq:CCSNrate}
f_{\rm CCSN} = \frac{\int_{0.1}^{100} M \psi(M)dM}{\int_{M_{\rm min}}^{M_{\rm max}}\psi(M)dM},
\end{equation}
where $0.1 M_\odot$ and $100 M_\odot$ in the numerator is the range of the IMF, and $M_{\rm min}$ to $M_{\rm max}$ in the denominator is the mass range of stars that explode as CCSNe. Due to the steep IMF slope, the majority of CCSNe occur near the minimum mass threshold, making $M_{\rm min}$ the most important parameter. Its value has been statistically determined by combining $20$ Type IIP supernova progenitor observations to be $M_{\rm min}^{\rm IIP} = 8.5^{+1}_{-1.5} M_\odot$ \citep{Smartt09}. This is consistent with the highest masses estimated for white dwarf progenitors, $\sim 7 \, {\rm M_\odot}$ \citep{Williams09}. Thus, two different approaches seem to be converging to $M_{\rm min} \approx 8 \pm 1 \, {\rm M_\odot}$. As we show below, this uncertainty affects $f_{\rm CCSN}$ by $\sim 20$\%.

The value of $M_{\rm max}$ is more uncertain than $M_{\rm min}$, but is less important. The maximum mass from Type IIP supernova progenitors is found to be $M_{\rm max}^{\rm IIP} = 16.5 \pm 1.5 \, M_\odot$ \citep{Smartt09}, but consideration of other types of SNe---including Types IIn and Ibc---yields larger values. For example, it is widely expected that Type Ibc SNe originate from evolved massive stars that have shed their envelopes, whose initial masses are $\gtrsim 25 \, M_\odot$. The spatial distribution of CCSNe supports that Type Ibc progenitors must be more massive than those of Type IIP SNe \citep{AJ08,AJ09}. Theoretically, stars with masses above $M_{\rm max} \sim 40 \, M_\odot$ may promptly form black holes whose optical transient phenomena remains uncertain \citep[e.g.,][]{Fryer99,Kochanek08}. We adopt $M_{\rm max} = 40 \, M_\odot$ as our canonical value, which together with $M_{\rm min} = 8 \pm 1 \, M_\odot$ yields $f_{\rm CCSN} = 147^{+27}_{-29} \, M_\odot^{-1}$, a $\sim 20$\% uncertainty. On the other hand, varying $M_{\rm max}$ between $25 M_\odot$ and $100 M_\odot$ only affects $f_{\rm CCSN}$ by $\sim 5$\%. We refer the reader to \cite{Horiuchi11} for detailed discussions of the uncertainties in the SFR-to-CCSN rate connection.

One of the advantages of CCSNe is that the SFR normalization can be probed without significant dependence on the IMF. This is because the mass range of stars exploding as CCSNe is similar to the mass range of stars powering the radiation used to measure the SFR. For example, we can write explicitly the SFR estimated from the UV and CCSNe as $f_{\rm UV} L_{\rm UV}$ and  $f_{\rm CCSN} \dot{N}_{\rm CCSN}$. Although changes to the IMF affects  $f_{\rm UV}$ and $ f_{\rm CCSN}$ individually by factors of $\sim 2$, their ratio remains fairly constant. For the SalA and BG IMFs, $ f_{\rm UV}/f_{\rm CCSN} = 1.09 \times 10^{-30} \, {\rm yr^{-1} \, erg^{-1}\,s\,Hz}$ and $1.01 \times 10^{-30} \, {\rm yr^{-1} \, erg^{-1}\,s\,Hz}$, respectively, which are within $10$\% of the Salpeter IMF value $1.09 \times 10^{-30} \, {\rm yr^{-1} \, erg^{-1}\,s\,Hz}$. 

\subsection{Comparison at Cosmic Distances}

Measurements of the CCSN rate at cosmic distances have rapidly increased in the past decade. Taking the measurements of \cite{Cappellaro99,Dahlen04,Cappellaro05,Botticella08,Bazin09,Li11}, the observed CCSN rate is well-fit by a strong redshift evolution of $(1+z)^{3.3}$ between $0<z<1$ with a $z=0$ normalization of $0.78^{+0.2}_{-0.2} \times 10^{-4} \, {\rm Mpc^{-3} \, yr^{-1}}$. By comparison, the CCSN rate predicted from the SFR using the NOROT calibration and Equation (\ref{eq:CCSNrate}) has a $z=0$ normalization that is twice as large, $1.6^{+0.3}_{-0.2} \times 10^{-4} \, {\rm Mpc^{-3} \, yr^{-1}}$. Uncertainties associated with calculating the CCSN rate from the SFR seem not to be able to satisfactorily explain the difference \citep[e.g.,][]{HB06,Horiuchi11}. 

Rotation helps bridge the normalization difference by decreasing the SFR and hence the predicted CCSN rate. However, the reduction is expected to be at most $\sim 30$\%, which is not enough to bridge the difference completely. Furthermore, it has been argued that a combination of intrinsically dim and heavily dust-obscured CCSNe being missed in cosmic CCSN surveys could be responsible for the observed discrepancy \citep{Horiuchi11}. The argument is based on the high fraction of dim CCSNe in the local $<10$ Mpc volume. \cite{Mattila12} further investigated the dim fraction in more detail, notably by including host galaxy extinction. Their results have been included in the most recent CCSN rate measurements by \cite{Dahlen12}  and \cite{Melinder12}, who find the CCSN rate in the range $z \sim 0.4$--$1.1$ to match the cosmic SFR at those redshifts. In the redshift range $0 < z <0.4$ the simple application of the local dim CCSN fraction has a similarly positive effect \citep{Horiuchi11,Mattila12}, although it remains to be confirmed by new CCSN rate measurements.

\subsection{Comparison at Local Distances} \label{sec:localCCSN}

Discoveries of CCSNe in the nearby volume have also improved recently. In Table \ref{table:local}, we show a list of CCSNe discovered within $\sim 11$ Mpc in the past $12$ years from $2000$ to $2011$ inclusive. Similar compilations can be found in \cite{Horiuchi11,Botticella12,Mattila12}. We exclude previous years due to known supernova incompleteness issues \citep[][]{HB10}. Distances are taken from the $11$HUGS catalog \citep{K08}. These are estimated from a combination of direct stellar distances (where available) and flow-corrected radial velocities. The Local Group flow model of \cite{KM96} is used, which the authors find to be more accurate for the $11$ Mpc volume than a Virgocentric flow model. The last three columns show whether the host galaxy is included in the respective galaxy catalogs as labeled.

Included in Table \ref{table:local} are SN~$2008$S-like events (SN~$2008$S, SN~$2002$bu, NGC$300$-OT, SN~$2010$da, and PSN J$12304185$+$4137498$) whose true nature continues to be debated. They are characterized by explosions that are $2$--$3$ mag dimmer than regular CCSNe, with narrow emission lines, and evidence of internal extinction; their progenitors are also dust enshrouded \citep{Prieto08}. While they are relatively common among nearby CCSNe, their progenitors are extremely rare \citep{Khan10}, leading \cite{Thompson09} to propose that many massive stars go through this dust-obscured phase shortly ($\sim 10^4$ yr) before the explosion. SN~$2008$S-like events may be true CCSNe, perhaps of the electron-capture type arising from AGB stars \citep{Thompson09,Botticella09,Pumo09}, or they may be extreme versions of non-explosive outbursts \citep{Thompson09,Smith09,Bond09,Berger09,Kashi10,Smith11}. For a comprehensive discussion of these and related models, see, e.g., \cite{Thompson09,Smith11,K11,Kochanek12}.
 
The compilation is a lower limit of the true CCSNe occurrence. This is because the CCSN discoveries are mainly made by surveys and amateurs with varying search strategies and methods, instead of a complete systematic survey. Although modern telescopes using CCDs are capable of discovering a typical CCSN within $11$ Mpc, CCSNe can be missed due to a combination of incomplete galaxy coverage, low survey cadence, heavily dust-obscured or dim CCSNe, or obstruction by the Galactic plane and the Sun. Indeed, the majority of CCSNe have occurred in large, well-known galaxies: all $22$ CCSNe occurred in the full $11$HUGS galaxy catalog. Also, $15$ of these occurred in our galaxy sample with LVL cuts. Since the LVL sky-coverage is only $51$\%, we are likely missing some CCSNe occurring in the remaining volume. Furthermore, the northern hemisphere is more closely observed, resulting in somewhat more CCSNe discovered with positive declinations than with negative declinations ($14$ versus $8$). As a specific example illustrating these points, SN~$2008$jb remained undiscovered by all galaxy-targeted supernova searches despite its close distance of $9.6$ Mpc. The host galaxy was a dwarf galaxy in the southern hemisphere, and the CCSN was only discovered in archival images of all-sky surveys \citep{Prieto12}.

There are $17$ definite CCSNe and $5$ SN~$2008$S-like events in Table \ref{table:local}. With our LVL sky-coverage cuts, this is reduced to $11$ definite CCSNe and $4$ SN~$2008$S-like events. From Poisson statistics, the $1 \sigma$ lower and upper limits of the $11$ definite CCSNe are $7.7$ and $15.4$, which gives a CCSN rate of  $1.8^{+0.7}_{-0.5} \, {\rm yr^{-1} }$ within the $11$ Mpc volume, where we have applied a sky-coverage correction analogous to the local SFR calculations. The SFR required is simply the CCSN rate multiplied by $f_{\rm CCSN}$. For our canonical $f_{\rm CCSN}$, this gives $270^{+110}_{-80} \, {\rm M_\odot \, yr^{-1}}$. Repeating the exercise including SN~$2008$S-like events gives a CCSN rate of $2.5^{+0.8}_{-0.6} \, {\rm yr^{-1} }$ and a required SFR of $360^{+120}_{-90} \, {\rm M_\odot \, yr^{-1}}$. 

The SFR values are summarized in Table \ref{table:results}. The SFR required by CCSNe is consistent given uncertainties with the cosmic SFR extrapolation and the non-rotating UV-SFR, but does not overlap within the uncertainties with all other SFR estimates. If SN~$2008$S-like events are included, there is a strong tension with all direct local SFR estimates. The H$\alpha$-SFR show a more significant tension with the SFR inferred from CCSNe, but as we have discussed, the CCSN rate probes SFR on time scales that are longer than the H$\alpha$ indicator and more comparable to the UV. 

The SFR values required by CCSNe can be reduced by adopting a larger mass range for CCSNe. However, even with extreme values of $M_{\rm min} = 7 M_\odot$ and $M_{\rm max} = 100 M_\odot$, the required SFR is about $200 \, {\rm M_\odot \, yr^{-1}}$ ($280 \, {\rm M_\odot \, yr^{-1}}$) excluding (including) SN~$2008$S-like events. One possible solution is to further decrease the minimum mass for CCSN. This was studied by \cite{Botticella12}, who suggested the CCSNe and non-rotating H$\alpha$-SFR are consistent if $M_{\rm min} = 6 M_\odot$. Although a similarly small $M_{\rm min}$ would be compatible with our rotating predictions, we do not favor this scenario because $M_{\rm min}$ has been accurately determined using progenitor observations of CCSNe in the nearby $\sim 20$ Mpc volume, i.e., a similar volume \citep{Smartt09}. Therefore, we treat these SFR estimates as the minimum SFR estimated from the observed CCSN counts.

To quantify whether the direct SFR estimates and CCSN counts are consistent, we assume a SFR estimate is true and calculate the probability of getting a test statistic at least as extreme as the number of CCSN observed. This is analogous to the determination of a $p$-value for an observation given a null hypothesis. We use our galaxy list with LVL cuts, i.e., $11$ confirmed CCSN discoveries in $12$ years. The non-rotating and rotating UV-SFR predict $6.9\pm1.0$ and $5.4\pm0.7$ CCSNe in the same list of galaxies over the same duration, respectively. Using Poisson statistics, the probability for at least $11$ CCSNe are $0.09$ and $0.02$, respectively. Therefore, neither non-rotating nor rotating models are good models for the observed CCSN counts, and the rotating model is considerably worse. The same probabilities for the H$\alpha$-SFR are $0.01$ and $5\times 10^{-4}$ for the non-rotating and rotating estimates, respectively. 
  
\begin{figure}[tb]
\centering\includegraphics[width=\linewidth,clip=true]{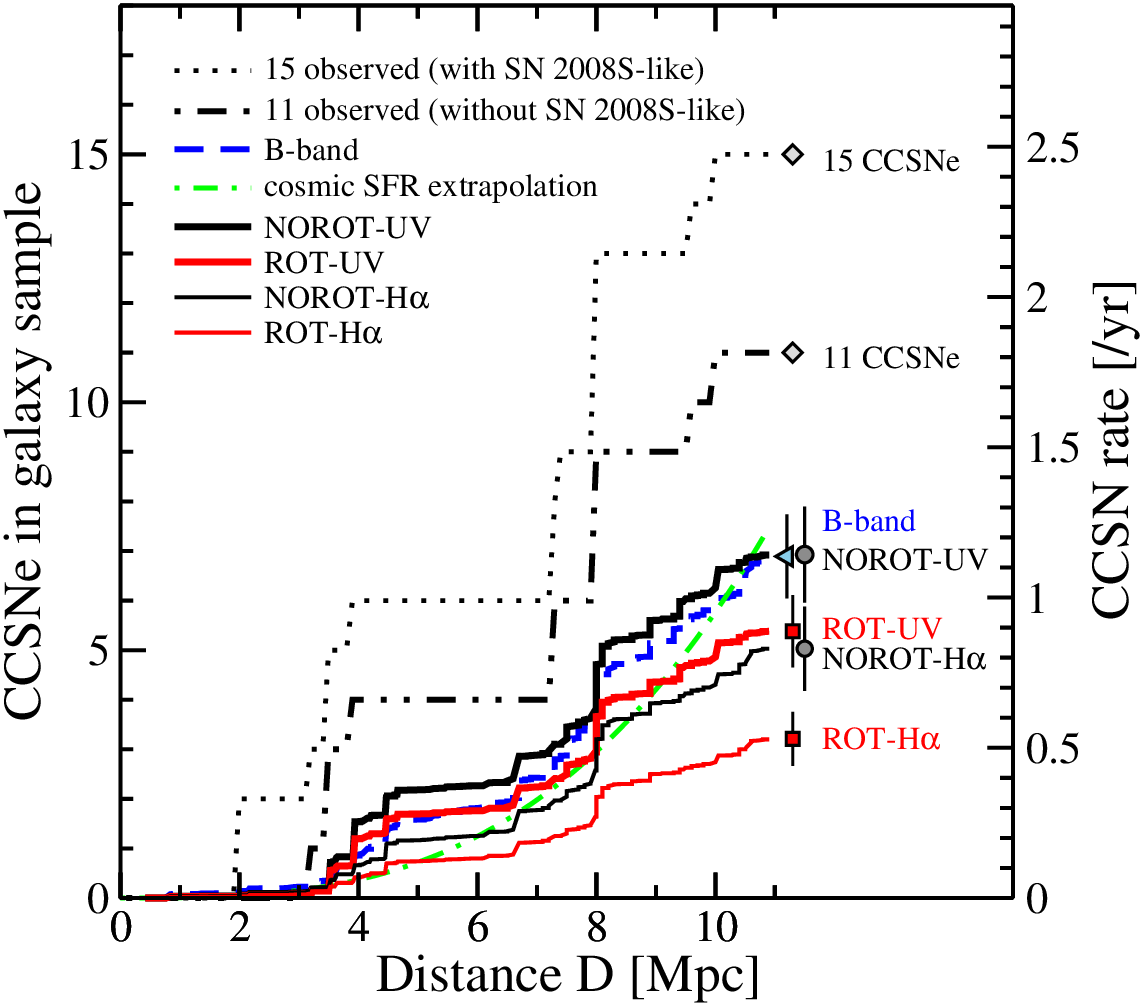}
\caption{Buildup (i.e., cumulative) of CCSN counts, showing the discovered CCSNe with (dotted) and without (dot-dot-dashed) SN~$2008$S-like events. Predictions from the SFR for H$\alpha$ (thin solid) and UV (thick solid) for non-rotating (black) and rotating (red) calibrations, and prediction based on the B-band (dashed), are also shown as labeled. All discoveries and predictions have been subject to the LVL angular cuts, i.e., a sky coverage of $51$\%. The total are shown by the data points on the right. Error bars are added to the model predictions, and show how there are more CCSNe discovered than predicted, in particular if the rotating calibration is assumed. On the right axis we convert counts to the {\it total} CCSN rate expected within distance $D$, i.e., the rates are calculated by correcting the CCSN counts and predictions upwards for the missing sky coverage (see the text). Note that the uncertainties on the CCSN rate estimates based on discovered CCSNe are not explicitly shown. Numerically, they are $1.8^{+0.7}_{-0.5} \, {\rm yr^{-1} }$ and $2.5^{+0.8}_{-0.6} \, {\rm yr^{-1} }$ without and with SN~$2008$S-like events, respectively. The former is marginally consistent with the NOROT-UV and B-band predictions, while the latter is significantly larger than all predictions (see also Table \ref{table:results}). \label{fig:localSN}}
\end{figure}	

In Figure \ref{fig:localSN}, we show the buildup (i.e., cumulative) of the CCSN counts in our galaxy list with LVL cuts, as a function of distance. The total within $11$ Mpc is shown as data points on the far right. On the right-hand $y$-axis, we convert the counts to a CCSN rate correcting for the sky-coverage. We show the observed CCSNe with and without SN~$2008$S-like events, and predictions based on the various SFR estimates, as labeled. We also show another CCSN estimate based on the galaxy absolute $B$-band luminosities using the latest SNu ($1 \, {\rm SNu} = 1 \, {\rm CCSN} \, /10^{10} L_{\odot,B} \, / {\rm century}$) of Lick Observatory Supernova Search \citep[LOSS;][]{Li11}. The SNu depends on the galaxy type as well as the galaxy $B$-band magnitude. The latter is the rate-size relation found by the LOSS group, which introduces a weighting such that smaller galaxies have higher SNu \citep[the LOSS rates without the rate-size relation are in good agreement with previous estimates by][]{Cappellaro99}. We use the same galaxy sample as we did for the SFR calculation. We correct the galaxy $B$-band luminosity for foreground Galactic extinction but not for host extinction, since the LOSS SNu have not been corrected for host extinction either. We then sum the CCSN rate from individual galaxies.

From Figure \ref{fig:localSN}, it is clear that observed CCSN counts are significantly larger than those predicted from the SFR. Also, it is interesting how the UV-SFR and $B$-band predictions are quantitatively very similar. Both predict $7 \pm1$ CCSNe in our galaxy sample over $12$ years, or a CCSN rate of $1.1 \pm 0.2 \, {\rm yr^{-1}}$ within $11$ Mpc. If this is interpreted as the robustness of CCSN rate predictions, it leads to the conclusion that there has been a chance occurrence of excess CCSNe around $\sim 4$ Mpc where the discovered rate first overshoots the predictions. The statistics are low in this range: only $4$ out of $11$ CCSNe are within $4$ Mpc. However, the $B$-band predicts only $0.85$ CCSNe, and the probability to get at least as extreme as the number of CCSN observed is $0.01$ (for the entire $11$HUGS sample, the respective numbers are $6$ CCSNe discovered, $1.3$ predicted, and $0.002$). So, it implies a rare occurrence. It is interesting that \cite{Mattila12} also find an excess of CCSNe at a similar distance of $\sim 6$ Mpc, based on matching the local CCSNe discoveries to the cosmic CCSN rate measurements. Note however that this also requires a rather unlikely upward fluctuation, since the predicted number of CCSNe in the entire $11$HUGS sample is $2.6$ while the observed is $8$. Finally, it should be noted that the cosmic CCSN rate predicted by \cite{Li11} is a factor $\sim 2$ smaller than the cosmic SFR rate. Although the local and cosmic rates cannot be directly compared, it leaves the possibility that the prediction in the local volume based on the $B$-band is also slightly low. 

\section{Discussions and Summary} \label{sec:summary}

Based on the new non-rotational (NOROT) and rotational (ROT) stellar tracks of \cite{E12}, we used the stellar evolutionary synthesis code PEGASE.2 to derive SFR calibration factors for non-rotating and rotating stellar populations. The ROT population shows significantly \emph{increased} photon output compared to the NOROT population, which leads to a \emph{decrease} in SFR calibration factors of $30$\% and $40$\% for the UV and H$\alpha$ indicators, respectively (Table \ref{table:calibration}). Compared to the widely-used non-rotating SFR calibrations of \cite{K98}, the ROT calibrations are $10$\% and $40$\% smaller, respectively. These changes to the calibration factors are comparable to (or larger than) the uncertainty on the best-fit cosmic SFR of HB06 that is $20$\%--$30$\% ($1\sigma$--$3\sigma$) at redshifts between $0<z<1$. Thus it is evident that systematic effects such as the SFR calibration factors will need to be understood better for future precision SFR estimates. 

We compare the SFR estimates to the measured CCSN rates to investigate whether the rotational calibrations fit with data. At cosmic distances, the best-fit cosmic SFR is too high compared to the measured CCSN rate data by a factor $\sim 2$. Stellar rotation will decrease the cosmic SFR but not enough to bridge the difference completely. Furthermore, a part of the SFR-CCSN normalization discrepancy is likely caused by CCSNe that are missed due to a combination of intrinsically dim and heavily dust obscured CCSNe \citep{Mannucci07,Horiuchi11,Mattila12}. Recent cosmic CCSN rate measurements in the range $z>0.4$ that include this correction report no discrepancy with the cosmic SFR data \citep{Melinder12,Dahlen12}. 

At distances of $\sim 10$ Mpc, several studies have found hints of an excess of CCSNe \citep[e.g.,][]{ABY05,Kistler11,Horiuchi11,Kistler12}, while \cite{Botticella12} performed an in-depth investigation and found that the UV-SFR is consistent with the observed CCSNe. We make new estimates of the SFR and CCSN rate in the local $11$ Mpc volume, update the discovered CCSNe, and by comparing the two, find support of an excess of discovered CCSNe (Figure \ref{fig:localSN}). Consistent with \cite{Botticella12}, we find that the total UV-SFR is marginally consistent with the observed CCSNe within uncertainties. However, upon closer inspection, we find that this requires a chance excess of CCSNe in the $4$ Mpc volume which is rather unlikely (a $0.2$\%--$0.9$\% chance probability). The H$\alpha$-SFR underpredicts the observed CCSN rate considerably. Stellar rotation, which decreases both the UV- and H$\alpha$-SFR estimates, exacerbates the situation and is therefore disfavored by current data in the local volume. We emphasize that this result is strengthened by the fact that the locally discovered CCSN counts can only increase as surveys probe more complete galaxy lists at greater cadence and sensitivities. The result is also robust to local overdensities, since both the SFR and CCSN rates will be similarly affected.

Thus, there is a dichotomy between the cosmic CCSN rate predicted from the SFR, which tends to overestimate the cosmic CCSN measurements, and the local CCSN rate predicted from the observed SFR, which underpredicts the local CCSN observations. Unless the SFR calibrations systematically change with distance, the emerging picture is that the SFR calibration factors are constrained to be close to or only slightly larger than the currently used values. If it were smaller, it would underpredict the local CCSN rate. If it were larger, it would severely overpredict the cosmic CCSN rate. However, it may be slightly larger, because not all massive stars need to explode as optically bright CCSNe, although the fraction of dark collapses is expected to not exceed $\sim 50$\% \citep[e.g.,][]{Kochanek08}. 

More generally, the local CCSNe discoveries set a lower bound on the SFR, and using this fact we disfavor any significant change to the total energy output among massive stars, including, but not limited to, rotation. An example is the impact of binaries. PEGASE.2 treats binarity only for the purposes of determining chemical enrichment by Type Ia supernovae, but in reality binary interactions affect various aspects of stellar evolution, including mixing, mass loss, and surface rotation. The binary fraction of massive stars is known to be at least in the several tens of percent range \citep[e.g.,][]{Mason98} and more recently up to $50$\% \citep{Sana13}. Additionally, a significant fraction of binaries are estimated to be in triple or higher order systems \citep{GM01}, which can lead to interesting interactions \citep[e.g.,][]{Thompson11}. In fact, binary interactions can spin stars up to the extent that may leave little room for single rapidly rotating stars \citep{deMink13}. Recently, \cite{Zhang12} found that binaries lead to a reduction of the SFR calibrations by $\sim 0.2$ dex; such changes are also constrained by the observed CCSNe, unless binaries change the mass thresholds for CCSNe.

Explanations of the CCSN-SFR dichotomy is itself of interest. Generally, effects that affect one distance regime and not the other are required. Given the simple counting and lower limit nature of the locally discovered CCSNe, the most obvious option in the local volume is to increase the SFR. However, increasing the SFR calibrations will necessarily increase the cosmic SFR as well. This leaves galaxy dust correction as the most realistic explanation. Another possibility is that the local galaxy distances are systematically underestimated. Since the SFR scales with the luminosity, and the CCSNe are simple counting, a $\sim 30$\% systematic increase in the galaxy distances will increase the SFR to match the discovered CCSN counts. For example, reducing the Hubble constant will increase the distances, but if we adopted $H_0=70 \, {\rm km \, s^{-1} \, Mpc^{-1}}$ this would only increase the distances by $\sim 7$\%, which is not enough. For the cosmic situation, the cosmic CCSN rate may be too low. One possibility is that the cosmic CCSN measurements are missing dim CCSNe as explained above. \cite{Horiuchi11} show that the fraction of dim CCSNe observed in the local volume is higher than at cosmic distances, and argue that incorporating dim CCSNe increases the cosmic CCSN measurements to match the cosmic SFR. \cite{Mattila12} include host galaxy extinction and find less dim CCSNe but nonetheless an important increase in the cosmic CCSN rates. Alternatively, the cosmic luminosity density used to derive many cosmic CCSN rates may be too low, although such a systematic change over a wide redshift range is not independently motivated.

Changes to the SFR will cause other effects not discussed in this paper. For example, they will affect predictions of the stellar mass density \citep[e.g.,][]{Madau98,MP00,Chary01,Cole01}. Current predictions based on the HB06 cosmic SFR are higher by a factor $\sim 2$ than the direct measurements \citep{HB06}, a fact that has been investigated in the context of IMF shape variations and IMF evolution \citep{WTH08,Wilkins08}. Rotation decreases the SFR and hence the stellar mass density prediction, but the comparison to data is more complex. This is because the stellar mass measurements are derived using galaxy spectral energy distribution fittings, and thus are also affected by stellar rotation. Since rotation increases the luminosity output of stars, it is likely to reduce the stellar mass density measurements also. Therefore, variations of the IMF may still be needed. However, quantitative predictions are beyond the scope of this paper since it requires calculating longer times including effects of low and intermediate-mass stars. The SFR has also been compared to the extragalactic background light \citep{HBD09,RM12,Inoue12}. However, this is not a strong probe of the SFR calibration variations, because a smaller calibration is effectively canceled by a larger energy output per stellar population. 

One of the important uncertainties is the unknown distribution of initial stellar rotation velocities. The rotational tracks of \cite{E12} assume a single initial rotational velocity proportional to the critical breakup velocity, $v_{\rm ini} = 0.4 v_{\rm crit}$, for all stellar masses. In reality, true stellar populations are expected to have a distribution of rotational velocities. \cite{Brott11} calculate stellar tracks for almost a dozen initial rotational velocities, and find that the effects of rotation are only appreciable after a threshold rotational velocity is reached. This includes the main-sequence lifetime increase that affects the SFR calibration factors. Thus, realistic distributions of rotation velocities will likely reduce the effects estimated in this paper. However, it should be mentioned that at the same time, the many benefits of the \cite{E12} rotational tracks (Section \ref{sec:introduction}) will also be negated, since they require rotation at the level assumed in \cite{E12} to be ubiquitous among stars.

Another uncertainty is the small number of CCSNe (between $11$ and $22$) and the small number of star-forming galaxies (the largest $10$ galaxies contribute $\sim 45$\% of the total SFR within $11$ Mpc) that make dust correction uncertain. Performing comparisons at larger distances with more CCSNe and galaxies will dramatically improve the situation since even an increase in factor of two in distance would give a factor $\sim 8$ increase in the rate. Presently running observation programs such as the All-Sky Automated Survey for the Brightest Supernovae \citep[ASAS-SN; e.g.,][]{Khan11} will find CCSNe in a volume-limited sample of nearby galaxies, and the Palomar Transient Factory will collect larger numbers of CCSNe in survey mode \citep{Law09,LienFields09}. Issues such as completeness should therefore improve dramatically in the near future. Uncertainties inherent to the SFR--CCSN comparison, e.g., the progenitor mass range for CCSN, is rapidly reducing \citep{Smartt09b}. CCSNe will therefore provide an excellent and compelling measure of the SFR in the future that would enable a new way of studying various SFR systematic effects. 


\acknowledgments
We thank Andy Gould, Andrew Hopkins, Jennifer Johnson, Rob Kennicutt, Kohta Murase, Marc Pinsonneault, Jose Prieto, Kris Stanek, and David Weinberg for useful discussions. We especially thank Christopher Kochanek for discussions and a careful reading of the manuscript. This research made use of the IAU Central Bureau for Astronomical Telegrams and the Sternberg Astronomical Institute supernova catalogs and the NASA/IPAC Extragalactic Database (NED), which is operated by JPL/Caltech, under contract with NASA. This work is supported in part by a JSPS fellowship for research abroad (to SH), and by NSF Grant PHY-1101216 (to JFB).



\end{document}